\documentclass{emulateapj}
\shortauthors{Toft et al.}
\shorttitle{The size - starformation relations at z=2}
\bibdata{letter}

\begin{document}

\title{The size-star formation relation of massive galaxies at $1.5<z<2.5$ }
\author{S. Toft\altaffilmark{1,2}, M. Franx\altaffilmark{3}, P. van Dokkum\altaffilmark{4}, N. M. F{\" o}rster Schreiber\altaffilmark{5}, I. Labbe\altaffilmark{6}  S.  Wuyts{\altaffilmark{7}}, { D. Marchesini \altaffilmark{8}} }



\altaffiltext{1}
{Dark Cosmology Centre, Niels Bohr Institute, University of Copenhagen, Juliane Mariesvej 30, DK-2100 Copenhagen, Denmark, email: sune@dark-cosmology.dk}

\altaffiltext{2}
{European Southern Observatory, D-85748 Garching, Germany}

\altaffiltext{3}
{Leiden Observatory, Postbus 9513, 2300 RA Leiden, Netherlands, email: franx@strw.leidenuniv.nl}

\altaffiltext{4}
{Department of Astronomy, Yale University, New Haven, CT 06520-8101, email: dokkum@astro.yale.edu, danilo.marchesini@yale.edu}

\altaffiltext{5}
{Max-Planck-Institut fuer extraterrestrische Physik, Postfach 1312, D-85748 Garching, Germany, email: forster@mpe.mpg.de}

\altaffiltext{6}
{Carnegie Observatories, 813 Santa Barbara street, Pasadena, CA-91101, email: ivo@ociw.edu; Hubble Fellow}

\altaffiltext{7}
{Harvard-Smithsonian Center for Astrophysics, 60 Garden street, Cambridge, MA 02138, email: swuyts@cfa.harvard.edu; W. M. Keck Postdoctoral Fellow }

\altaffiltext{8}
{Tufts University, email: danilo.marchesini@tufts.edu }

\begin{abstract}
We study the relation between size and star formation activity in a complete sample of 225 massive ($M_*>5\cdot10^{10} M_{\odot}$) galaxies at $1.5<z<2.5$, selected from the FIREWORKS UV-IR catalog of the CDFS. Based on stellar population synthesis model fits to the observed restframe UV-NIR SEDs, and independent MIPS 24$\mu m$ observations, 65\% of galaxies are actively forming stars, while 35\% are quiescent. Using sizes derived from 2D surface brightness profile fits to high resolution ($FWHM_{PSF}\sim 0.45\arcsec)$ groundbased ISAAC data, we confirm and improve the significance of the relation between star formation activity and compactness found in previous studies, using a large, complete mass-limited sample. At $z\sim2$, massive quiescent galaxies are significantly smaller than massive star forming galaxies, and a median factor of $0.34\pm0.02$ smaller than galaxies of similar mass in the local universe. 13\% of the quiescent galaxies are unresolved in the ISAAC data, corresponding to sizes $<1 kpc$, more than 5 times smaller than galaxies of similar mass locally. 
The quiescent galaxies span a Kormendy relation which, compared to the relation for local early types, is shifted to smaller sizes and brighter surface brightnesses and is incompatible with passive evolution.
The progenitors of the quiescent galaxies, were likely dominated by highly concentrated, intense nuclear star bursts at $z\sim 3-4$, in contrast to star forming galaxies at $z\sim2$ which are extended and dominated by distributed star formation.

\end{abstract}
\vspace{2 cm}
\section{Introduction}

The size of a galaxy is a important property, which scales with its mass.  The  evolution of galaxy sizes with redshifts provide strong constraints on galaxy  formation models, as it reflects the change in the distribution of stellar mass with time. Sizes of high redshift galaxies provide the strongest constraints.  Lately evidence has been accumulating for massive high redshift galaxies having  significantly smaller sizes than local galaxies of similar mass \citep{toft2005, daddi2005,trujillo2006b, trujillo2006a, trujillo2007, toft2007, zirm2007, cimatti2008, vandokkum2008a, franx2008}. There is furthermore  evidence for a relation between star formation activity and size: at $z\sim2$,  star forming galaxies 
are on average a factor of 2 smaller than local  galaxies of similar mass, while quiescent galaxies on average are smaller by a  factor of 3-6 \citep{toft2007,zirm2007,cimatti2008,vandokkum2008a}, suggesting the existence of a significant population of massive  galaxies with extremely dense old stellar populations, which are not easily explained by current galaxy formation models.  
All these observations are based  on relatively small samples of galaxies, with a range of possible biases  resulting from different selection schemes. In this paper we  present the first study focusing on the correlation between star formation activity and size in a large mass-limited sample of galaxies at $z\sim 2$.
 
Throughout the paper we assume a flat cosmology with $\Omega_m=0.3$, $\Omega_\Lambda=0.7$ and  $h=0.72$. Magnitudes are in the AB system.

\section{FIREWORKS} 
This paper is based on FIREWORKS, an imaging survey of $\sim$ 138 arcmin$^2$ in the CDFS, which combines deep multi waveband space and ground based optical-NIR data with MIR data from the Spitzer IRAC and MIPS instruments \citep{wuyts2008}. We adopt the photometric redshifts of \cite{wuyts2008} which were derived from the UV-IR SEDs (calibrated using 1477 spectroscopic redshifts) and have an accuracy of $\Delta z/(1+z) \sim 0.05$ at $z_{spec}>1$, 
and the SED modeling of \cite{forster2009} who fits \cite{bruzual} stellar population synthesis models to derive constraints on the star formation rates, histories, ages, stellar masses, extinction etc. We here use the results for a Calzetti extinction law, solar metallicity and a Salpeter IMF, \citep[which we renormalized to a \cite{kroupa2001} IMF by multiplying derived masses and star formation rates by $10^{-0.2}$, see ][]{franx2008}. We note that the derived masses decrease on average by a factor of 1.4 when \cite{maraston2005} models are used. There is no evidence for this factor being dependent on color, mass or redshift \citep{wuyts2007}.
We constructed a complete \citep[see ][]{franx2008}, mass-limited ($M>5\cdot10^{10}M_{\odot}$) sample of 225 galaxies in the photometric redshift range $1.5\le z_{phot}\le 2.5$, which form the basis of the following analysis.
More details of the fits and accuracies of the derived parameters are presented in \cite{wuyts2008} and \cite{forster2009}.

\begin{figure*}
\plottwo{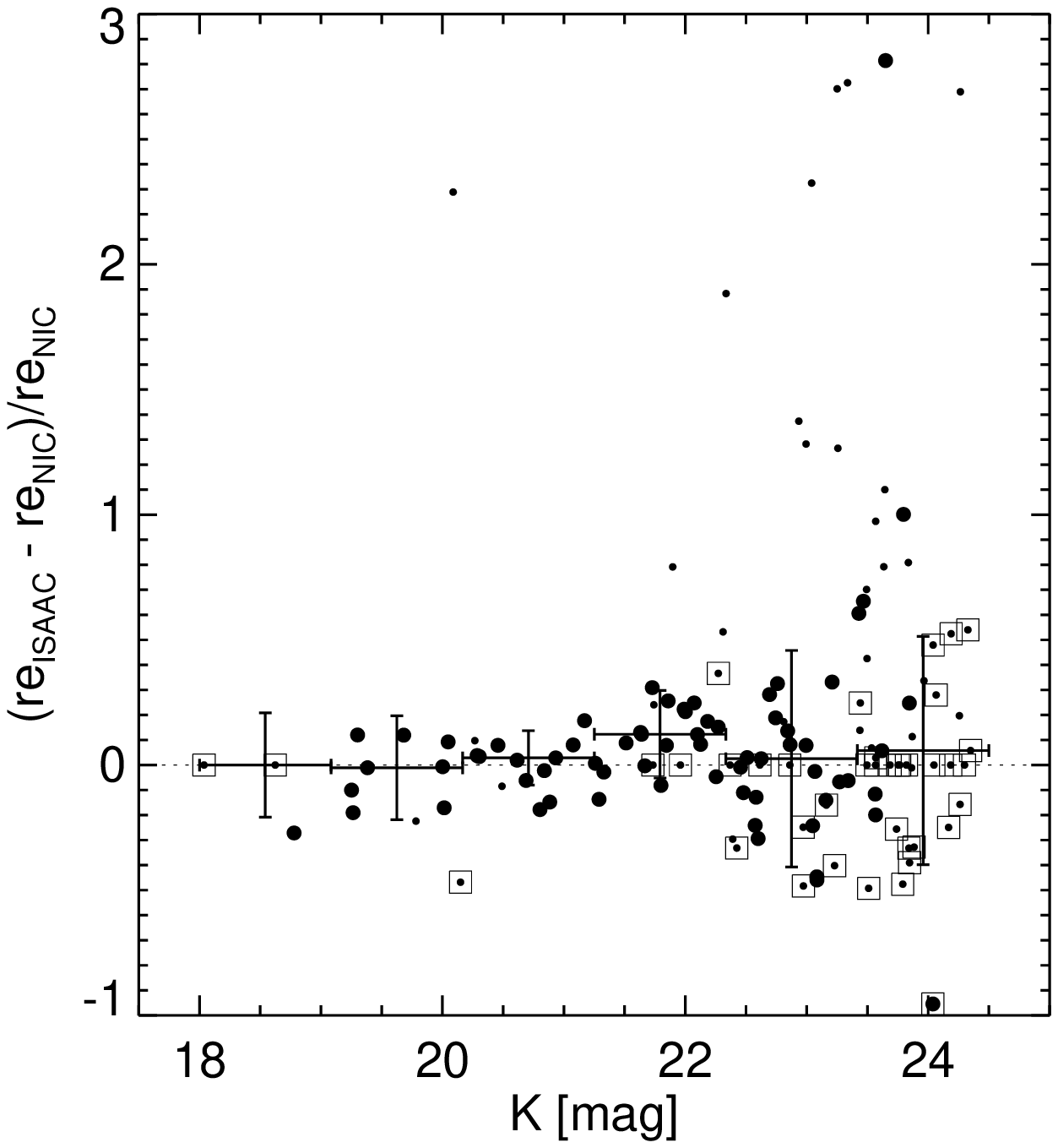}{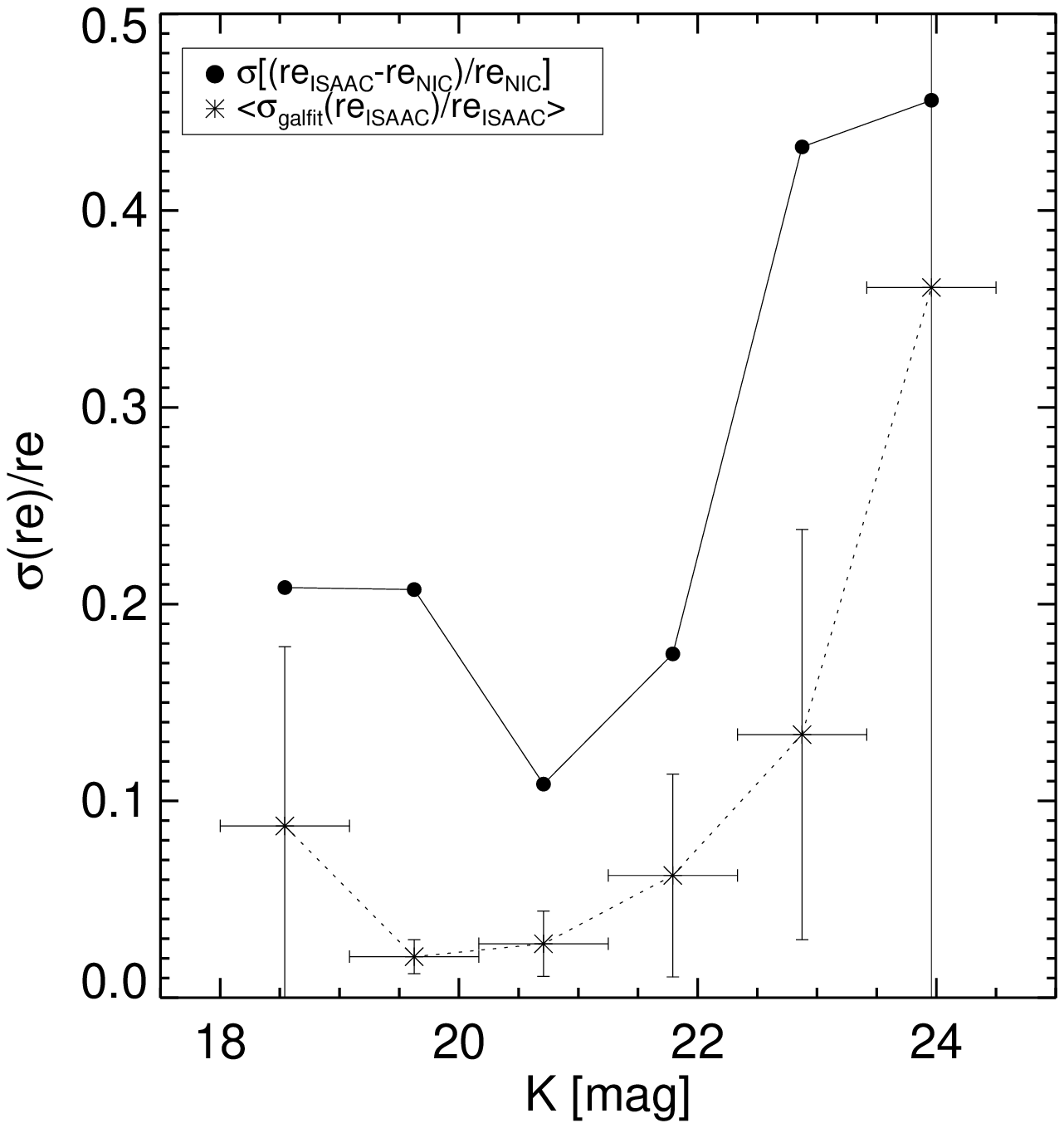}
\caption{{{\bf Left}: Comparison of sizes derived from ISAAC K band and NICMOS/NIC3 F160W band data, as a function of K-band magnitude. Small/large points are galaxies $re_{NIC}$ smaller/larger than $FWHM_{NIC3}/2$ respectively.  Squares indicate galaxies which are unresolved in the ISAAC data. There is no systematic difference, but the scatter increases with magnitude.  {\bf Right:} Relative $1\sigma$ scatter as a function of K-band magnitude (full curve), derived from Fig. \ref{niccompare}a. The scatter is 10-20\% at $K<22$, and increases at fainter magnitudes to $\sim 45\%$ at $K=24$. Also plotted is the median fitting uncertainty (and scatter) reported by {\sc galfit} for the ISAAC K band image (dotted curve).
  }}
\label{niccompare}
\end{figure*}

\section{Size fitting}
We fitted the 2D surface brightness distribution of the galaxies in the ISAAC K band images (pixel scale $0.147\arcsec/pix$, seeing FWHM$\sim 0.45\arcsec$), using {\sc galfit} \citep{peng2002} and a \cite{sersic1968} profile. We allowed the Sersic index to vary between $1\le n\le 4$, and derived the ``circularized'' effective radius $re$, enclosing half the light of the model, in a similar way as in \cite{toft2007}.
Objects with derived $re< 0.5$ pixel are considered unresolved and assigned an upper limit on their $re$ corresponding to 0.5 pixel. As an independent check we derived sizes of a number of stars in the field, using other stars as PSFs. All the stars had derived sizes $\ll$ 0.5 pixel. We verified that no systematic uncertainties in the derived sizes arise from the size of the fitting box, or the choice of PSF star.
The CDFS has been observed with HST/ACS as part of the GOODS-S survey \citep{giavalisco2004}, and part of it has been observed with the HST/NIC3 camera, as part of the Hubble ultra deep field project \citep{thompson2005}. 
In Fig \ref{niccompare} we compare the sizes of 126 galaxies with overlapping ISAAC K-band and NIC3 F160W-band data.
The NIC3 data is  more than three magnitudes deeper than the ISAAC data (10 $\sigma$ limiting magnitude $F160W=27.8$), and has better resolution $FWHM_{NIC3}\sim0.35\arcsec$, so with this comparison we can test the derived sizes for systematic uncertainties introduced by limited resolution and signal to noise. 
As can be seen from Fig. \ref{niccompare}a, there is no systematic difference between the sizes derived from ISAAC and NICMOS. There is a significant scatter which increases with magnitude. This is shown in Fig. \ref{niccompare}b,  where we plot the relative $1\sigma$ error derived from Fig. \ref{niccompare}a as a function of magnitude. The scatter is 10-20\% at $K<22$, and increases at fainter magnitudes to $\sim 45\%$ at $K=24$. Also plotted is the median fitting uncertainty (and scatter) reported by ({\sc galfit} for the ISAAC K band image. Interestingly the {\sc galfit} errors trace the empirically determined trend with magnitude, but systematically underestimates its value.
A small fraction (14\%) of the galaxies have $re_{ISAAC}>>re_{NIC}$. These are  faint ($K\gtrsim 23$) and small ($re_{NIC}< FWHM_{NIC}/2$).  The sizes derived from the ISAAC data for the smallest faintest galaxies should thus be considered upper limits.

Using a fitting procedure identical to the one used here, \cite{trujillo2006a} derived sizes from the FIRES ISAAC data \citep{labbe2003, forster2006} which is of similar depth and quality, and was acquired and reduced in an identical way to the data set studied here. They studied in great details the effects of low signal to noise, and small intrinsic sizes on the derived structural parameters, using simulations,  and found no significant systematic biases in the derived sizes, but an increasing scatter with magnitude, in agreement with the conclusion from Fig.\ref{niccompare}.

We also fitted the sizes of the galaxies in our sample in the ACS F606W band images. Based on these fits, we removed 16 galaxies from the sample which had bright blue central sources with derived sizes consistent with being unresolved, since their broad band photometry and derived sizes are likely to be contaminated by active galactic nuclei (AGN).

\section{Results}

Based on the specific star formation rates (sSFR) derived from the SED fits, ($65\%$) of galaxies in the sample has $sSFR>0.03\rm{Gyr}^{-1}$ and are classified as ``star forming''  (median $\left < sSFR \right>_{sf}=1.13\pm0.67 ~Gyr^{-1}$), while $35\%$ has $sSFR<0.03 Gyr^{-1}$ and are classified as ``quiescent'' (median $\left < sSFR \right>_{q}=0.01\pm 0.01~Gyr^{-1}$).   
These fractions are similar to those found for distant red galaxies \citep{kriek2006, toft2007, zirm2007} as expected, since most massive galaxies are red at these redshifts.  The corresponding volume densities are $\phi_q=1.7^{+0.6}_{-0.5}\cdot10^{-4}\rm{Mpc}^{-3}$ and $\phi_{sf}=3.1_{-0.8}^{+1.1}\cdot10^{-4}\rm{Mpc}^{-3}$ for quiescent and star forming galaxies respectively \citep[error bars represent the estimated cosmic variance, see][]{marchesini2008}. 
The results are robust with respect to small changes in the sSFR threshold used for dividing the galaxies into star forming and quiescent, as relatively few galaxies ($\lesssim 10\%$) have intermediate ($0.03 Gyr^{-1}<sSFR<0.5 Gyr^{-1}$) derived specific star formation rates. 
 
\begin{figure}
\epsscale{1.1}
\plotone{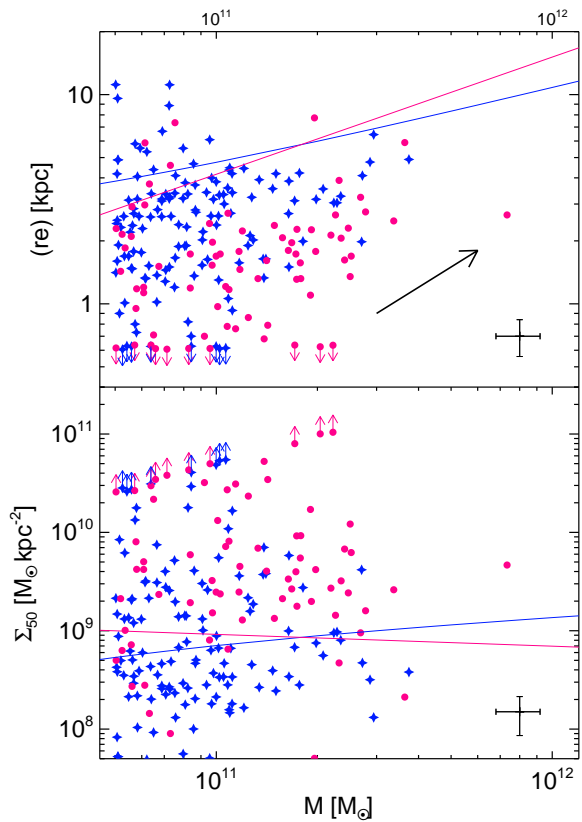}
\caption{{{\bf Top}: Mass-size relation for $M_* >5\cdot10^{10}M_{\odot}$ galaxies with $1.5<z_{phot}<2.5$. Filled red circles are quiescent galaxies, blue stars are star forming. The curves represent the mass-size relation of early type (red) and late-type (blue) galaxies in SDSS \citep{shen2003}. Quiescent galaxies are significantly smaller than star forming galaxies of similar mass. The quiescent and  star forming galaxies are median factors of $0.34\pm 0.02$ and $0.51\pm0.02$ smaller than galaxies of similar mass locally. The arrow shows the increase in mass and size predicted from equal mass dry merger simulations \citep{boylan-kolchin2006}. {\bf Bottom:} Surface mass density versus stellar mass. Star forming galaxies are a median factor of $4.1\pm0.2$ denser than galaxies of similar mass locally, while quiescent galaxies are $8.8 \pm 0.7$ times denser.  Crosses show typical error bars for galaxies∂ with median properties}}
\label{masssize} 
\end{figure}

In Fig. \ref{masssize} we plot the mass of the galaxies versus their effective radii. At a given mass, the quiescent galaxies are significantly smaller than the star forming galaxies. With the exception of a few dozen galaxies with $M<10^{11}M_{\odot}$, which within the scatter are consistent with the local relation, the galaxies are smaller than local SDSS galaxies of similar mass at all considered masses.

We parametrize the evolution of the mass-size relation with $log~re(M_*)_{z=2}=log~re(M_*) + logA$, where A is the median offset between the relation at $z=2$ and $z=0$, and use monte carlo simulations to derive the most likely value of A and its uncertainty. We generate 500 realizations of the data set by varying the photometric redshifts and corresponding stellar masses randomly within their $1\sigma$ errors (derived from monte carlo simulations by \cite{wuyts2008}). For each photo-z realization we calculate the corresponding physical size in kpc, and randomly perturb it according to the $re$ uncertainties shown in Fig.\ref{niccompare}, derived from {\sc galfit} and from the comparison to NICMOS (added in quadrature). For each realization we then calculate the median A, and estimate its uncertainty from its variation in the realizations. In this way we find that
the star forming galaxies are a median factor of  $A_{sf}=\left<re/re_{SDSS}\right>_{sf}=0.51 \pm 0.02$ smaller than galaxies of similar mass in the local Universe, while the quiescent galaxies are smaller by a median factor of $A_{q}=\left<re/re_{SDSS}\right >_{q} = 0.34 \pm 0.02$. This difference is smaller than was found for quiescent DRGs \citep[$\left <re_{DRG}/re_{SDSS} \right >\sim 0.2$, ][]{toft2007} and K-selected galaxies without emission lines at $z\sim 2.3$, \citep[$\left <re_{DRG}/re_{SDSS} \right >\sim 0.17$][]{vandokkum2008a}, but the scatter is large, with values ranging from $0.1\lesssim re_{q}/re_{SDSS}\lesssim 1$.  This difference is likely a consequence of the systematically higher redshifts ($2<z<3.5$) and brighter K-band magnitudes of the galaxies in these studies relative to the median galaxy in the present mass selected sample.
 $13\%$ of the quiescent galaxies are unresolved in the ISAAC data with $re$ set to an upper limit of 0.5 pix. Furthermore, from Fig.\ref{niccompare} we concluded that the sizes of some of the smallest, faintest galaxies may be overestimated form the ISAAC data.  These galaxies are included in the calculation of $A_q$, which should thus be considered a lower limit for the size difference. 
Also plotted in  Fig.\ref{masssize} is the surface mass density $\Sigma_{50}=\frac{M_*/2}{\pi re^2}$ of the galaxies as a function of stellar mass. 
From the monte carlo simulations described above, we find that the quiescent galaxies are a median factor of $\left< \Sigma_{50}/\Sigma_{50, SDSS}\right >_q=8.8 \pm 0.7$ denser than galaxies of similar mass locally,  while the star forming galaxies are denser by a median factor of $\left< \Sigma_{50}/\Sigma_{50, SDSS}\right >_{sf}=4.1\pm0.2$.

In the above size comparison, we compared to the local late-type relation when the best fitting sersic $n$ was $<2.5$ and the local early type relation when $n\ge2.5$, consistent with the definition of the relations in \cite{shen2003}. One may worry that this could introduce systematic effects since the fitted n parameter is relatively uncertain, especially for the smallest galaxies, and we do not know for sure if the $z\sim2$ galaxies will evolve into early or late type galaxies, however since the local relations are very similar in the considered mass range, the derived median offsets from the local mass-size relation does not change significantly when the sizes, independently of the fitted $n$, are compared to only the local early type ($0.35\pm 0.02$) or late type ($0.33\pm0.02$) relations.

\begin{figure}
\epsscale{1.2}
\plotone{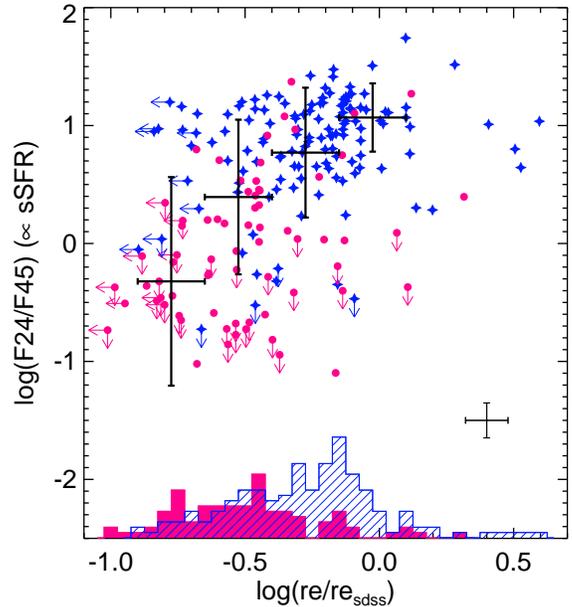}
\caption{{Ratio of flux at $24\micron$ and $4.5\micron$ versus size (normalized by the size of similar mass galaxies in SDSS). $F_{24}/F_{4.5}$ scales with the specific star formation rate, since $F_{24}$ scales with star formation activity (and redshift) and $F_{4.5}$ scales with stellar mass \citep[See][] {toft2007}. Symbols are as in Fig 1. The histograms represent the size distribution of the quiescent (full, red) and star forming (blue, hatched) galaxies.  Thick crosses represent running medians of the distributions (vertical  bars: stddev around median, horizontal: bin size) and clearly confirms the correlation between star formation activity and size.  Most of the quiescent galaxies has low $F_{24}/F_{4.5}$ ratios (many of them are undetected at 24$\micron$), while most of the star forming galaxies have $log(F_{24}/F_{4.5}) \sim 1$.} {The thin cross in the lower right corner show typical error bars for a galaxy with median properties.}}
\label{mips}
\end{figure}

As an independent confirmation of the correlation we added MIPS 24$\micron$ imaging to the analysis. As shown by \cite{toft2007}, the ratio of flux at 24 and 4.5$\micron$ as measured by the Spitzer MIPS and IRAC instruments is a good empirical tracer of the specific star formation rate (and/or AGN activity) of galaxies in this redshift range. In Fig. \ref{mips} we plot this ratio ($F24/F4.5$) as a function of size (normalized by the local mass size relation). The majority of the quiescent galaxies have $log(F24/F4.5) \lesssim 0$ (many of them only have upper limit detections, corresponding to $sSFR\lesssim 0.2 Gyr^{-1}$) while most of the star forming galaxies have $log(F24/F4.5) \sim 1$, corresponding to $sSFR \sim 1.5 Gyr^{-1}$, independently confirming the correlation between size and star formation activity (based on SED fits), and that the quiescent galaxies are small and red due to extremely compact old quiescent stellar populations rather than dust enshrouded star formation or AGN activity.

The relation between star formation activity and size found here for a mass selected sample at $z\sim2$ is consistent with the relation between between color, size and mass and its evolution with redshift found by \cite{franx2008} for a K-selected sample. At a given mass they found galaxy sizes to evolve like $re \propto 1/(1+z)^{0.59\pm0.10}$, corresponding to a size difference of $re_{z=2}/re_{z=0}=0.52\pm0.06$, which is very similar to the size difference measured here for the whole sample (quiescent+star forming galaxies): $re_{z=2}/re_{SDSS}=0.48 \pm 0.03$. For quiescent galaxies they found a faster evolution $re \propto 1/(1+z)^{1.09\pm0.07}$, corresponding to $re_{z=2}/re_{z=0}=0.30\pm0.03$, again consistent with the value derived here for quiescent galaxies $re_{z=2}/re_{SDSS}=0.34\pm 0.02$.
    
\begin{figure}
\epsscale{1.2}
\plotone{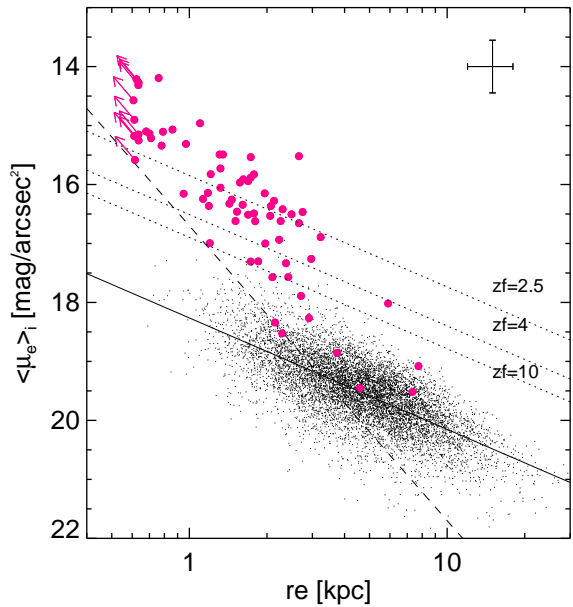}
\caption{{Kormendy relation: Restframe i-band surface brightness (within $re$, corrected for dust extinction) versus size for quiescent galaxies in the sample. Black dots represent early type galaxies in SDSS. The cross represent the typical error bars of a quiescent galaxy with median properties.  The dashed line represent the approximate completeness limit, due to the mass selection. The dotted lines represent predicted Kormendy relations at $z=2$ for galaxies that formed at $z_f=2.5, 4$ and 10 and passively evolve onto the local relation \citep{bruzual}. The observed relation is inconsistent with passive evolution. The smallest galaxies have too high surface brightnesses, to be consistent, even with the $z_f=2.5$ prediction, and passive evolution to $z\sim 0$ would result in a local population of small bright massive galaxies which are not observed.   }}

\label{kormendy}
\end{figure}

In Fig.\ref{kormendy} we plot the Kormendy relation between size and dust corrected restframe i-band surface brightness for the quiescent galaxies. Also plotted are the Kormendy relation of early type galaxies in SDSS, and the predictions of simple passive evolution models which, if the stellar populations form at a common formation redshift $z_f$, just shifts the relation vertically without changing its slope (if the galaxies formed at different redshifts, the passive evolution could also change the slope). The $z\sim2$ quiescent galaxies are shifted to smaller sizes and higher surface brightnesses, than locally.
The $z\sim 2$ relation appear to be steeper, but this may be due to selection effects, as indicated by the dashed line, which shows the approximate position of galaxies at the mass limit ($5\cdot10^{10}M_{\odot}$).   
Using bootstrapping, we find that the surface brightness of the $z\sim2$ galaxies on median are $3.2\pm0.1\,mag/arcsec^2$ brighter than  similar size local galaxies. This is consistent  with results from of smaller samples at similar redshifts \citep{toft2007,longhetti2007, cimatti2008, vandokkum2008a}.
Note that a population of small ($re <1 kpc$) quiescent elliptical galaxies exist in the local universe (e.g in the Virgo cluster, see \cite{ferrarese2006}), which is missed by SDSS selection function. However these are typically faint, low mass dwarf galaxies, with surface brightnesses in agreement with the local Kormendy relation shown in Fig.\ref{kormendy}.   

The $z\sim2$ relation can not evolve into the local relation through simple passive evolution, as this would result in galaxies which are too small and bright for their mass, compared to what is observed. 
In other words, the galaxies need to evolve in size in addition to the passive fading of their stellar light to evolve into the local population.

\section{Discussion}
A possible formation scenario for the ultra compact quiescent galaxies is through high redshift gas rich mergers, which models suggests under certain circumstances can create very dense stellar populations \citep[see e.g.][]{khockfar2006}. 
Examples of high redshift gas rich mergers with intense associated star formation are sub-mm selected galaxies (SMGs), which in some cases have very high central concentrations of molecular gas and dynamical mass densities comparable to the central stellar density found here in the quiescent galaxies \citep[e.g.][]{tacconi2008, coppin2009}. For this reason, an evolutionary connection between the two galaxy types has been hypothesized \citep{toft2007, cimatti2008, tacconi2008}.  
A simple connection is however unlikely. 
In contrast to the quiescent galaxies, SMGs are among the most vigorously star forming galaxies known in the universe. 
High extinction and evidence for a large fraction of interacting/merging systems among SMGs complicate the comparison of their sizes with those of more common star forming galaxies at $z\sim2$.  \cite{smail2004} and \cite{almaini2005} find half light radii of a few to $\sim$ 10 kpc for samples of $\sim10$ SMGs, which are more similar to the range spanned by our star forming sub sample, and thus consistent with the SFR-size trend (but see \cite{tacconi2008} for a contrasting view).
Such large inferred sizes, together with extended disturbed morphologies and dynamics seen in a fraction of SMGs, are clearly different than for many of the quiescent galaxies studied here.  It seems implausible that all SMGs could be
 turned off and appear as compact remnants within very short timescales (the redshift distributions are similar).
The old stellar ages measured for the quiescent galaxies \citep[0.5-1.5 Gyr, ][]{kriek2006}, and their compact structure suggest that they formed the majority of their stars in highly concentrated (nuclear) star bursts at $3\lesssim z \lesssim5$. 
This suggests that a population of very compact, intensely star forming galaxies exists at $z\sim4$, with properties quite different from the bulk of star forming galaxies at $z\sim2$, which are extended, and dominated by spatially distributed star formation.
If their star formation is relatively unobscured, these galaxies may be present in B-band and V-band drop out selected samples of $z\sim 4$ and $z\sim5$ galaxies. Interestingly, the space densities of the brightest ($L_{UV}> 2.5 L_{UV}^*$) B-band and  ($L_{UV}> 2 L_{UV}^*$) V-band drop out galaxies are similar to the space density of quiescent $z\sim2$ galaxies \citep{bouwens2007},  however their (restframe UV) sizes are considerable larger \citep[$\left< re \right >= 1.6 kpc$,][]{bouwens2004}.
If their star formation is triggered by gas rich major mergers, it may be obscured by dust, in which case these galaxies would not be present in Lyman drop selected samples, but may be detectable as a population of dense  $z\sim4$ sub-mm galaxies. 
Identifying SMGs at these redshifts is however difficult, since they fall below the detection limit of the deepest radio VLA maps, needed for identifying their optical counterpart due to the poor resolution of current sub-mm facilities \citep[e.g. ][]{ivison2007}.
Therefore, only a few examples of SMGs are known at these redshifts  \citep[e.g.][]{younger2007, knudsen2008, dannerbauer2008}, all of which have been detected in radio. However, a significant fraction of SMGs \citep[about 1/3, ][]{ivison2007} are  unidentified in radio and are thus likely to be at $z>3.5$. Some of these may be progenitors of the $z\sim2$ quiescent galaxies.

While the derived stellar mass densities in the quiescent galaxies ($10^{9}-10^{11}M_{\odot}kpc^{-3}$) are much higher than in local massive galaxies, comparable or higher densities are found in globular clusters and ultra compact dwarf galaxies \citep[$10^{10}-10^{13} M_{\odot}kpc^{-3}$,][]{dabringhausen2008}, which are some of the oldest stellar systems known. Even though their total masses are much smaller ($10^6-10^8M_{\odot}$), and there may not be a direct connection to the compact $z\sim 2$ quiescent galaxies, it is interesting to note that star bursts in the early Universe can result in extremely compact stellar distributions.

The quiescent galaxies at $z\sim 2$ are dominated by old extremely compact stellar populations, so it would be natural to assume that they would continue to evolve passively into local elliptical galaxies, which are the oldest, densest, most massive galaxies in the local Universe.   
From Fig. \ref{masssize} and \ref{kormendy} it is clear that this scenario is ruled out, as passive evolution would result in too many extremely compact massive galaxies with too high surface brightnesses, compared to what is observed.
Some elliptical galaxies in the local universe have super solar metallicities.  Assuming super solar ($2.5Z_{\odot}$) metallicity in the SED fits, on average leads to $19 \pm1\%$ smaller masses for the quiescent galaxies, which is not enough to explain their unusually high mass densities.  

The high masses and small sizes of the quiescent galaxies corresponds to unusually high expected velocity dispersions ($\sigma=300-500 km/s$, see also \cite{toft2007,vandokkum2008a, cimatti2008}.  This has been observationally confirmed from a deep restframe optical spectrum of a $M=2*10^11 M_{\sun}$,  $z=2.2$ compact, quiescent galaxy which has a measured $\sigma=510^{165}_{-95} km/s$ \citep{vandokkum2009}. 
Recently, a small number of low redshift galaxies with velocity dispersions $\sigma_v > 350 km/s$, high stellar masses ($>10^{11}M_{\odot}$) and unusually small sizes (1-2 kpc) were found in the SDSS \citep{bernardi2006}. Their derived stellar surface mass densities are however not as extreme as for the quiescent galaxies found here \citep[see also][]{cimatti2008}, and they are extremely rare, with $10^3$ times smaller space densities ($\phi \sim 10^{-7}-10^{-8} Mpc^{-3}$), so a simple evolutionary connection is unlikely.

One of the most promising processes for decreasing the stellar mass density of the quiescent galaxies is dry merging. Simulations suggests that under favorable conditions, an equal mass dry merger can increase the size of the remnant by up to a factor of two \citep[e.g.][]{boylan-kolchin2006, naab2007, hopkins2008}. As indicated in Fig. \ref{masssize}, the compact quiescent galaxies would need to go through 2-3 successive equal mass mergers to end up near the local mass-size relation, which would result in remnant masses in excess of $5\cdot10^{11}M_{\odot}$.  
The space density of $M>5\cdot10^{11}M_{\odot}$ early type galaxies in the SDSS sample of \cite{bernardi2003} is $\sim 0.2\cdot10^{-4} Mpc^{-3}$, a factor of 10 lower than the density of the $z\sim2$ quiescent $M>5 \cdot10^{10}M_{\odot}$ galaxies considered here. Interestingly, three successive equal mass dry mergers of the quiescent $z\sim2$ galaxies could thus result in remnants with masses and number densities comparable to those of the most massive local elliptical galaxies, but possibly not with with quite as large sizes \citep[see also e.g.][]{bezanson2009}. 

Recent studies have explored minor dry merging as an efficient way of decreasing the central stellar mass density of the $z\sim 2$ quiescent galaxies to the value observed in local ellipticals, and the possibility that the compact stellar populations of some of them could survive to the present day, hidden in centres of local ellipticals underneath lower density envelopes of stars accreted at later times \citep{naab2009, bezanson2009, hopkins2009}. Finally, if the IMF evolves with time and were more top heavy in the past as suggested by e.g. \cite{vandokkum2008b}, 
part of the offset from the local mass-size relation could be explained by a systematic over estimation of the masses at $z\sim2$.

We thank Thomas R. Greve and Steffen Mieske and Mariska Kriek for useful discussions, and the anonymous referee, for useful suggestions which improved the analysis. S. Toft gratefully acknowledges support from the Lundbeck Foundation.  S. Wuyts gratefully acknowledges support from the W. M. Keck Foundation.


\begin{thebibliography}{35}
\expandafter\ifx\csname natexlab\endcsname\relax\def\natexlab#1{#1}\fi

\bibitem[{{Almaini} {et~al.}(2005){Almaini}, {Dunlop}, {Conselice}, {Targett},
  \& {Mclure}}]{almaini2005}
{Almaini}, O., {Dunlop}, J.~S., {Conselice}, C.~J., {Targett}, T.~A., \&
  {Mclure}, R.~J. 2005, astro-ph/0511009

\bibitem[Bezanson et al.(2009)]{bezanson2009} Bezanson, R., van 
Dokkum, P.~G., Tal, T., Marchesini, D., Kriek, M., Franx, M., 
\& Coppi, P.\ 2009, \apj, 697, 1290 

\bibitem[{{Bernardi} {et~al.}(2003){Bernardi}, {Sheth}, {Annis}, {Burles},
  {Eisenstein}, {Finkbeiner}, {Hogg}, {Lupton}, {Schlegel}, {SubbaRao},
  {Bahcall}, {Blakeslee}, {Brinkmann}, {Castander}, {Connolly}, {Csabai},
  {Doi}, {Fukugita}, {Frieman}, {Heckman}, {Hennessy}, {Ivezi{\'c}}, {Knapp},
  {Lamb}, {McKay}, {Munn}, {Nichol}, {Okamura}, {Schneider}, {Thakar}, \&
  {York}}]{bernardi2003}
{Bernardi}, M. et al., 2003,
  \aj, 125, 1817

\bibitem[{{Bernardi} {et~al.}(2006){Bernardi}, {Sheth}, {Nichol}, {Miller},
  {Schlegel}, {Frieman}, {Schneider}, {Subbarao}, {York}, \&
  {Brinkmann}}]{bernardi2006}
{Bernardi}, M. et al., 2006, \aj, 131, 2018

\bibitem[{{Bouwens} {et~al.}(2004){Bouwens}, {Illingworth}, {Blakeslee},
  {Broadhurst}, \& {Franx}}]{bouwens2004}
{Bouwens}, R.~J., {Illingworth}, G.~D., {Blakeslee}, J.~P., {Broadhurst},
  T.~J., \& {Franx}, M. 2004, \apjl, 611, L1

\bibitem[{{Bouwens} {et~al.}(2007){Bouwens}, {Illingworth}, {Franx}, \&
  {Ford}}]{bouwens2007}
{Bouwens}, R.~J., {Illingworth}, G.~D., {Franx}, M., \& {Ford}, H. 2007, \apj,
  670, 928

\bibitem[{{Boylan-Kolchin} {et~al.}(2006){Boylan-Kolchin}, {Ma}, \&
  {Quataert}}]{boylan-kolchin2006}
{Boylan-Kolchin}, M., {Ma}, C.-P., \& {Quataert}, E. 2006, \mnras, 369, 1081


\bibitem[{{Bruzual} \& {Charlot}(2003)}]{bruzual}
{Bruzual}, G. \& {Charlot}, S. 2003, \apj, 405, 538

\bibitem[{{Cimatti} {et~al.}(2008){Cimatti}, {Cassata}, {Pozzetti}, {Kurk},
  {Mignoli}, {Renzini}, {Daddi}, {Bolzonella}, {Brusa}, {Rodighiero},
  {Dickinson}, {Franceschini}, {Zamorani}, {Berta}, {Rosati}, \&
  {Halliday}}]{cimatti2008}
  {Cimatti}, A. et al., 2008, \aap, 482, 21

\bibitem[Coppin et al.(2009)]{coppin2009} Coppin, K.~E.~K., et 
al.\ 2009, \mnras, 395, 1905 



\bibitem[{{Dabringhausen} {et~al.}(2008){Dabringhausen}, {Hilker}, \&
  {Kroupa}}]{dabringhausen2008}
{Dabringhausen}, J., {Hilker}, M., \& {Kroupa}, P. 2008, \mnras, 386, 864

\bibitem[{{Daddi} {et~al.}(2005){Daddi}, {Dickinson}, {Chary}, {Pope},
  {Morrison}, {Alexander}, {Bauer}, {Brandt}, {Giavalisco}, {Ferguson}, {Lee},
  {Lehmer}, {Papovich}, \& {Renzini}}]{daddi2005}
{Daddi}, E. et al., 2005, \apjl, 631, L13

\bibitem[{{Dannerbauer} {et~al.}(2008){Dannerbauer}, {Walter}, \&
  {Morrison}}]{dannerbauer2008}
{Dannerbauer}, H., {Walter}, F., \& {Morrison}, G. 2008, \apjl, 673, L127

\bibitem[Ferrarese et al.(2006)]{ferrarese2006} Ferrarese, L., et 
al.\ 2006, \apjs, 164, 334 

\bibitem[F{\"o}rster Schreiber et al.(2006)]{forster2006} 
F{\"o}rster Schreiber, N.~M., et al.\ 2006, \aj, 131, 1891 


\bibitem[{{F{\" o}rster Schreiber} {et~al.}(2009){F{\" o}rster Schreiber},
  {Franx}, {van Dokkmum}, {Labbe}, {Wuyts}, \& {Toft}}]{forster2009}
{F{\" o}rster Schreiber}, N.~M. et al., 2009, \apj, in prep

\bibitem[Franx et al.(2008)]{franx2008} Franx, M., van Dokkum, 
P.~G., Schreiber, N.~M.~F., Wuyts, S., Labb{\'e}, I., 
\& Toft, S.\ 2008, \apj, 688, 770 

\bibitem[{{Giavalisco} {et~al.}(2004){Giavalisco}, {Ferguson}, {Koekemoer},
  {Dickinson}, {Alexander}, {Bauer}, {Bergeron}, {Biagetti}, {Brandt},
  {Casertano}, {Cesarsky}, {Chatzichristou}, {Conselice}, {Cristiani}, {Da
  Costa}, {Dahlen}, {de Mello}, {Eisenhardt}, {Erben}, {Fall}, {Fassnacht},
  {Fosbury}, {Fruchter}, {Gardner}, {Grogin}, {Hook}, {Hornschemeier}, {Idzi},
  {Jogee}, {Kretchmer}, {Laidler}, {Lee}, {Livio}, {Lucas}, {Madau},
  {Mobasher}, {Moustakas}, {Nonino}, {Padovani}, {Papovich}, {Park},
  {Ravindranath}, {Renzini}, {Richardson}, {Riess}, {Rosati}, {Schirmer},
  {Schreier}, {Somerville}, {Spinrad}, {Stern}, {Stiavelli}, {Strolger},
  {Urry}, {Vandame}, {Williams}, \& {Wolf}}]{giavalisco2004}
{Giavalisco}, M. et al., 2004, \apjl, 600, L93

\bibitem[{Hopkins} {et~al.}(2008)]{hopkins2008}
{Hopkins}, P. et al., 2008, \apj (submitted)  

\bibitem[Hopkins et al.(2009)]{hopkins2009} Hopkins, P.~F., Bundy, 
K., Murray, N., Quataert, E., Lauer, T.~R., 
\& Ma, C.-P.\ 2009, \mnras, 1043 


\bibitem[{{Ivison} {et~al.}(2007){Ivison}, {Greve}, {Dunlop}, {Peacock},
  {Egami}, {Smail}, {Ibar}, {van Kampen}, {Aretxaga}, {Babbedge}, {Biggs},
  {Blain}, {Chapman}, {Clements}, {Coppin}, {Farrah}, {Halpern}, {Hughes},
  {Jarvis}, {Jenness}, {Jones}, {Mortier}, {Oliver}, {Papovich},
  {P{\'e}rez-Gonz{\'a}lez}, {Pope}, {Rawlings}, {Rieke}, {Rowan-Robinson},
  {Savage}, {Scott}, {Seigar}, {Serjeant}, {Simpson}, {Stevens}, {Vaccari},
  {Wagg}, \& {Willott}}]{ivison2007}
{Ivison}, R.~J. et al., 2007, \mnras, 380, 199

\bibitem[{{Khochfar} \& {Silk}(2006)}]{khockfar2006}
{Khochfar}, S. \& {Silk}, J. 2006, \apjl, 648, L21

\bibitem[{{Knudsen} {et~al.}(2008){Knudsen}, {Kneib}, \& {Egami}}]{knudsen2008}
{Knudsen}, K.~K., {Kneib}, J.-P., \& {Egami}, E. 2008, in Astronomical Society
  of the Pacific Conference Series, Vol. 381, Infrared Diagnostics of Galaxy
  Evolution, ed. R.-R. {Chary}, H.~I. {Teplitz}, \& K.~{Sheth}, 372

\bibitem[{{Kriek} {et~al.}(2006){Kriek}, {van Dokkum}, {Franx}, {Quadri},
  {Gawiser}, {Herrera}, {Illingworth}, {Labb{\'e}}, {Lira}, {Marchesini},
  {Rix}, {Rudnick}, {Taylor}, {Toft}, {Urry}, \& {Wuyts}}]{kriek2006}
{Kriek}, M. et al., 2006, \apjl, 649, L71

\bibitem[{{Kroupa}(2001)}]{kroupa2001}
{Kroupa}, P. 2001, \mnras, 322, 231

\bibitem[Labb{\'e} et al.(2003)]{labbe2003}  Labb{\'e}, I., et al.\ 2003, \aj, 125, 1107

\bibitem[{{Longhetti} {et~al.}(2007){Longhetti}, {Saracco}, {Severgnini},
  {Della Ceca}, {Mannucci}, {Bender}, {Drory}, {Feulner}, \&
  {Hopp}}]{longhetti2007}
{Longhetti}, M. et al., 2007, \mnras,
  374, 614

\bibitem[{{Maraston}(2005)}]{maraston2005}
{Maraston}, C. 2005, \mnras, 362, 799

\bibitem[Marchesini et al.(2008)]{marchesini2008} Marchesini, D., van 
Dokkum, P.~G., F{\"o}rster Schreiber, N.~M., Franx, M., Labb{\'e}, I., 
\& Wuyts, S.\ 2009, \apj, 701, 1765 



\bibitem[{Naab} {et~al.}(2007)]{naab2007} Naab, T., Johansson, 
P.~H., Ostriker, J.~P., \& Efstathiou, G.\ 2007, \apj, 658, 710

\bibitem[Naab et al.(2009)]{naab2009} Naab, T., Johansson, 
P.~H., \& Ostriker, J.~P.\ 2009, \apjl, 699, L178 


\bibitem[{{Peng} {et~al.}(2002){Peng}, {Ho}, {Impey}, \& {Rix}}]{peng2002}
{Peng}, C.~Y., {Ho}, L.~C., {Impey}, C.~D., \& {Rix}, H. 2002, \aj, 124, 266

\bibitem[{{S\'ersic}(1968)}]{sersic1968}
{S\'ersic}, J.~L. 1968, Atlas de Galaxias Australes (Cordoba: Obs. Astron.)


\bibitem[{{Shen} {et~al.}(2003){Shen}, {Mo}, {White}, {Blanton}, {Kauffmann},
  {Voges}, {Brinkmann}, \& {Csabai}}]{shen2003}
{Shen}, S., {Mo}, H.~J., {White}, S.~D.~M., {Blanton}, M.~R., {Kauffmann}, G.,
  {Voges}, W., {Brinkmann}, J., \& {Csabai}, I. 2003, \mnras, 343, 978

\bibitem[{{Smail} {et~al.}(2004){Smail}, {Chapman}, {Blain}, \&
  {Ivison}}]{smail2004}
{Smail}, I., {Chapman}, S.~C., {Blain}, A.~W., \& {Ivison}, R.~J. 2004, \apj,
  616, 71

\bibitem[{{Tacconi} {et~al.}(2008){Tacconi}, {Genzel}, {Smail}, {Neri},
  {Chapman}, {Ivison}, {Blain}, {Cox}, {Omont}, {Bertoldi}, {Greve}, {Foerster
  Schreiber}, {Genel}, {Lutz}, {Swinbank}, {Shapley}, {Erb}, {Cimatti},
  {Daddi}, \& {Baker}}]{tacconi2008} Tacconi, L.~J., et al.\ 
2008, \apj, 680, 246 



\bibitem[{{Thompson} {et~al.}(2005){Thompson}, {Illingworth}, {Bouwens},
  {Dickinson}, {Eisenstein}, {Fan}, {Franx}, {Riess}, {Rieke}, {Schneider},
  {Stobie}, {Toft}, \& {van Dokkum}}]{thompson2005}
{Thompson}, R.~I. et al., 2005, \aj,
  130, 1


\bibitem[{{Toft} {et~al.}(2005){Toft}, {van Dokkum}, {Franx}, {Thompson},
  {Illingworth}, {Bouwens}, \& {Kriek}}]{toft2005}
{Toft}, S., {van Dokkum}, P., {Franx}, M., {Thompson}, R.~I., {Illingworth},
  G.~D., {Bouwens}, R.~J., \& {Kriek}, M. 2005, \apjl, 624, L9

\bibitem[{{Toft} {et~al.}(2007){Toft}, {van Dokkum}, {Franx}, {Labbe},
  {F{\"o}rster Schreiber}, {Wuyts}, {Webb}, {Rudnick}, {Zirm}, {Kriek}, {van
  der Werf}, {Blakeslee}, {Illingworth}, {Rix}, {Papovich}, \&
  {Moorwood}}]{toft2007}
{Toft}, S. et al., 2007, \apj, 671, 285


\bibitem[{{Trujillo} {et~al.}(2006{\natexlab{a}}){Trujillo}, {Feulner},
  {Goranova}, {Hopp}, {Longhetti}, {Saracco}, {Bender}, {Braito}, {Della Ceca},
  {Drory}, {Mannucci}, \& {Severgnini}}]{trujillo2006b}
{Trujillo}, I. et al., 2006{\natexlab{a}}, \mnras, 373, L36

\bibitem[{{Trujillo} {et~al.}(2006{\natexlab{b}}){Trujillo}, {F{\"o}rster
  Schreiber}, {Rudnick}, {Barden}, {Franx}, {Rix}, {Caldwell}, {McIntosh},
  {Toft}, {H{\"a}ussler}, {Zirm}, {van Dokkum}, {Labb{\'e}}, {Moorwood},
  {R{\"o}ttgering}, {van der Wel}, {van der Werf}, \& {van
  Starkenburg}}]{trujillo2006a}
{Trujillo}, I. et al., 2006{\natexlab{b}}, \apj, 650, 18

\bibitem[{{Trujillo} {et~al.}(2007){Trujillo}, {Conselice}, {Bundy}, {Cooper},
  {Eisenhardt}, \& {Ellis}}]{trujillo2007}
{Trujillo}, I., {Conselice}, C.~J., {Bundy}, K., {Cooper}, M.~C., {Eisenhardt},
  P., \& {Ellis}, R.~S. 2007, \mnras, 382, 109

\bibitem[van Dokkum(2008)]{vandokkum2008b} van Dokkum, P.~G.\ 2008, 
\apj, 674, 29 

\bibitem[{{van Dokkum} {et~al.}(2008){van Dokkum}, {Franx}, {Kriek}, {Holden},
  {Illingworth}, {Magee}, {Bouwens}, {Marchesini}, {Quadri}, {Rudnick},
  {Taylor}, \& {Toft}}]{vandokkum2008a}
{van Dokkum}, P.~G. et al., 2008, \apjl, 677, L5

\bibitem[van Dokkum et al.(2009)]{vandokkum2009} van Dokkum, P.~G., 
Kriek, M., \& Franx, M.\ 2009, \nat, 460, 717 


\bibitem[Wuyts et al.(2007)]{wuyts2007} Wuyts, S., et al.\ 2007, 
\apj, 655, 51

\bibitem[{{Wuyts} {et~al.}(2008){Wuyts}, {Labbe}, {F{\" o}rster Schreiber},
  {Franx}, {Rudnick}, {Brammer}, \& {van Dokkum}}]{wuyts2008}
{Wuyts}, S., {Labbe}, I., {F{\" o}rster Schreiber}, N.~M., {Franx}, M.,
  {Rudnick}, G., {Brammer}, G.~B., \& {van Dokkum}, P.~G. 2008, \apj, 682, 985

\bibitem[{{Younger} {et~al.}(2007){Younger}, {Fazio}, {Huang}, {Yun}, {Wilson},
  {Ashby}, {Gurwell}, {Lai}, {Peck}, {Petitpas}, {Wilner}, {Iono}, {Kohno},
  {Kawabe}, {Hughes}, {Aretxaga}, {Webb}, {Mart{\'{\i}}nez-Sansigre}, {Kim},
  {Scott}, {Austermann}, {Perera}, {Lowenthal}, {Schinnerer}, \& {Smol{\v
  c}i{\'c}}}]{younger2007}
{Younger}, J.~D. et al., 2007, \apj, 671, 1531

\bibitem[{{Zirm} {et~al.}(2007){Zirm}, {van der Wel}, {Franx}, {Labb{\'e}},
  {Trujillo}, {van Dokkum}, {Toft}, {Daddi}, {Rudnick}, {Rix},
  {R{\"o}ttgering}, \& {van der Werf}}]{zirm2007}
{Zirm}, A.~W. et al., 2007, \apj, 656, 66

\end{thebibliography}
\end{document}